\documentclass[aps,prl,reprint,letterpaper,amsmath,amssymb,superscriptaddress,longbibliography]{revtex4-1}

\usepackage{color}

\newcommand{\delV}[1]{\textcolor{red}{\bf }}

\usepackage{graphicx}

\begin{document}
\title{Anomalous magnetic suppression of spin relaxation in a two-dimensional electron gas in a GaAs/AlGaAs quantum well}

\author{V.~V.~Belykh}
\email[]{belykh@lebedev.ru}
\affiliation{Experimentelle Physik 2, Technische Universit\"{a}t Dortmund, D-44221 Dortmund, Germany}
\affiliation{P. N. Lebedev Physical Institute of the Russian Academy of Sciences, 119991 Moscow, Russia}
\author{M.~V.~Kochiev}
\affiliation{P. N. Lebedev Physical Institute of the Russian Academy of Sciences, 119991 Moscow, Russia}
\author{D.~N.~Sob'yanin}
\email[]{sobyanin@lpi.ru}
\affiliation{P. N. Lebedev Physical Institute of the Russian Academy of Sciences, 119991 Moscow, Russia}
\author{D.~R.~Yakovlev}
\affiliation{Experimentelle Physik 2, Technische Universit\"{a}t Dortmund, D-44221 Dortmund, Germany}
\affiliation{Ioffe Institute, Russian Academy of Sciences, 194021 St. Petersburg, Russia}
\author{M.~Bayer}
\affiliation{Experimentelle Physik 2, Technische Universit\"{a}t Dortmund, D-44221 Dortmund, Germany}
\affiliation{Ioffe Institute, Russian Academy of Sciences, 194021 St. Petersburg, Russia}

\date{\today}
\begin{abstract}
We study the spin dynamics in a high-mobility two-dimensional electron gas confined in a GaAs/AlGaAs quantum well. An unusual magnetic field dependence of the spin relaxation is found: as the magnetic field becomes stronger, the spin relaxation time first increases quadratically but then changes to a linear dependence, before it eventually becomes oscillatory, whereby the longitudinal and transverse times reach maximal values at even and odd filling Landau level factors, respectively. We show that the suppression of spin relaxation is due to the effect of electron gyration on the spin-orbit field, while the oscillations correspond to oscillations of the density of states appearing at low temperatures and high magnetic fields. The transition from quadratic to linear dependence can be related to a transition from classical to Bohm diffusion and reflects an anomalous behavior of the two-dimensional electron gas analogous to that observed in magnetized plasmas.
\end{abstract}
\maketitle

\section{Introduction}
The electron spin relaxation in semiconductors without inversion center, such as GaAs, at low temperatures is usually governed by the random effective
magnetic fields acting on the electron spin \cite{Meier1984,Dyakonov2017}. These are the nuclear Overhauser field related to the hyperfine interaction of
the electron and nuclear spins \cite{Overhauser1953,Lampel1968} and the spin-orbit Dresselhaus or Rashba field arising from electron motion
\cite{Dresselhaus1955,BychkovRashba1984,Vasko1979}. In general, the following rule applies: for localized electrons the nuclear field governs the spin
relaxation, while for free electrons the spin-orbit effects are dominant.

In \emph{quantum dots}, where electrons are localized by three-dimensional confinement, the electron spin relaxation is determined by the nuclear environment \cite{Merkulov2002,Greilich2007,Chekhovich2013,Jaschke2017,Evers2018}. In \emph{bulk} semiconductors at low temperatures the degree of electron localization is determined by the donor concentration \cite{Dzhioev2002}. At low density of donors, the electrons are localized and the spin relaxation is due to the nuclear fields. On the other hand, at high doping density the electrons are free and the spin-orbit interaction governs the spin relaxation which is partly suppressed by collisions with donors (Dyakonov-Perel relaxation mechanism \cite{DyakonovPerel1971}). 
In high magnetic fields the situation becomes even more involved. In low-donor-density samples, the electron spin relaxation through nuclear spins becomes suppressed and spin diffusion comes into play \cite{Belykh2017}. In high-donor-density samples, where the Dyakonov-Perel mechanism is dominant, the spin relaxation is governed by pecularities of the electron motion, which are usually revealed in transport experiments. In particular, the longitudinal spin relaxation time $T_1$ is affected by the magnetic field similarly to the resistivity. With increasing magnetic field both $T_1$ and resistivity first decrease due to the suppression of weak localization \cite{Belykh2018} and then increase as a result of cyclotron precession \cite{Ivchenko1973}.

Of special interest is the case of \emph{quantum wells} (QWs) with a high mobility two-dimensional electron gas (2DEG). Here, the electrons are free and the spin relaxation is dominated by the Dyakonov-Perel mechanism. Note, that unlike for bulk systems, in QWs electron scattering events are rare since the donors are located in the barrier layers, so that the spin relaxation time is relatively short at zero field. With increasing magnetic field one can expect a strong manifestation of motional effects in the spin relaxation. The most well-known phenomenon related to the electron motion in 2DEG in magnetic field is the Quantum Hall (QH) effect \cite{Klitzing1980}. It takes place at low temperatures and high magnetic fields applied along the sample normal, when the separation between Landau levels exceeds the thermal broadening and inhomogeniety of the QW potential. Indeed, in works of Fukuoka et al. \cite{Fukuoka2008,Fukuoka2010} and Larionov et al. \cite{Larionov2015,Larionov2017,Larionov2020}, a sharp increase in the transverse inhomogeneous relaxation time $T_2^*$ was observed for magnetic fields corresponding to the odd filling factors $\nu = 2\pi \hbar n_\text{e} / eB_\text{F}$, where $n_\text{e}$ is the electron density, $B_\text{F}$ is the magnetic field along the sample normal (applied in Faraday geometry) and $e$ is the electron charge. This effect was explained in terms of skyrmions \cite{Fukuoka2010} and Goldstone mode formation \cite{Larionov2017,Larionov2020,Dickmann2019}. Furthermore, at $\nu = 2$ an extremely long longitudinal spin relaxation time $T_1$ was predicted \cite{Dickmann2013} and this fact was used for creation and manipulation of cyclotron magnetoexcitons \cite{Kulik2016,Kulik2018,Kulik2019}.

In this work, we investigate the longitudinal and transverse spin dynamics in a high-mobility 2DEG in a GaAs/AlGaAs QW taking advantage of the extended pump-probe Kerr rotation spectroscopy \cite{Belykh2016}. When increasing the magnetic field component along the QW normal, we observe an increase in the spin relaxation time with a rate inversely proportional to temperature. Here, the extreme relation $T_2^* \approx 2 T_1$ holds. However, at higher fields and low temperatures, the above relation breaks down and oscillations in $T_1$ and $T_2^*$ appear, so that $T_1$ has maxima at even filling factors, while $T_2^*$ has maxima at odd filling factors. To explain our experimental findings, we develop a theory that considers the combined action of spin precession in an effective spin-orbit field and electron cyclotron gyration in an external magnetic field. In the framework of this theory we show that the problem of spin relaxation can be reduced to the problem of electron spatial diffusion. The comparison of theory and experiment reveals an unexpected anomalous behavior of the high-mobility 2DEG that corresponds to Bohm diffusion and is analogous to the behavior observed under certain conditions in usual magnetized plasmas.

\section{Experimental details}
The results are obtained on a structure with a single modulation-doped GaAs QW containing a 2DEG with concentration $n_\text{e}$ of about $1 \times 10^{11}$ cm$^{-2}$. The structure is grown on a (001)-GaAs substrate followed by a thick GaAs buffer layer, a GaAs/Al$_{0.3}$Ga$_{0.7}$As
superlattice with 100 periods for strain relaxation, then by the 25-nm-wide quantum well of GaAs, followed by a thick Al$_{0.3}$Ga$_{0.7}$As layer
with Si $\delta$-doping, and the GaAs cap layer. The Hall mobility of the 2DEG in the QW $\mu_\text{e} > 2 \times 10^6$ cm$^2$/Vs at $T=2$~K.

For optical measurements the sample is placed in the variable temperature insert of a split-coil magnetocryostat ($T = 2-300$~K). The extended pump-probe Kerr rotation technique is used to study the electron spin dynamics. It is a modification of the standard pump-probe Kerr/Faraday rotation technique, where circularly-polarized pump pulses generate carrier spin polarization, which is then probed by the Kerr rotation of linearly-polarized probe pulses after reflection from the sample. Implementation of pulse picking for both pump and probe beams in combination with a mechanical delay line allows us to scan microsecond time ranges with picosecond time resolution. Details of the technique are given in Ref.~\cite{Belykh2016}. Here, a Ti:Sapphire laser emits a train of 2~ps pulses with a repetition rate of 76~MHz (repetition period $T_\text{R}=13.1$~ns). The pump protocol uses a single pulse per excitation period. The separation between these pulses is $80 T_\text{R} = 1050$~ns in order to clearly exceed the characteristic time of spin polarization decay. In most experiments, instead of the Kerr rotation, we measure the Kerr ellipticity, which is the ellipticity of the polarization of the initially linearly polarized probe beam after its reflection from the sample. The Kerr ellipticity, similarly to the Kerr rotation, is proportional to the spin polarization \cite{Yugova2009}, but depends less sensitively on the excitation energy. For measurements of the longitudinal spin relaxation time $T_1$, magnetic fields $B_\text{F}$ up to 6~T are applied parallel to the light propagation direction, that is, parallel to the sample growth ($z$-)axis (Faraday geometry). To measure the transverse spin relaxation time $T_2^*$, while maintaining the energy spectrum discrete, the sample is tilted so that its normal is $45^\circ$ to the light beams and to the magnetic field.

\section{Experimental results}
\begin{figure}
\includegraphics[width=1\columnwidth]{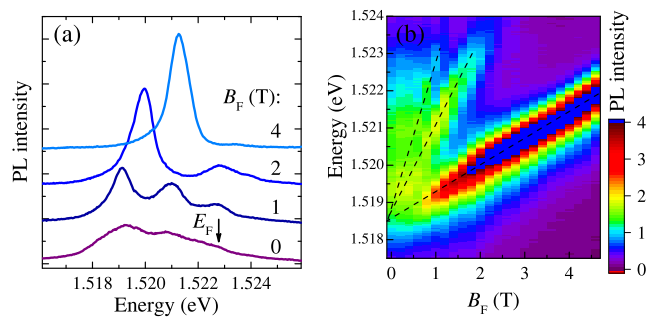}
\caption{(a) Photoluminescence spectra of a 2DEG for different magnetic fields $B_\text{F}$ applied in Faraday geometry for pulsed excitation with 1.771~eV central photon energy. Arrow marks the position of the Fermi energy at $B_\text{F} = 0$. (b) Corresponding two-dimensional energy-magnetic field map of the PL. Dashed lines show the positions of the Landau levels. Temperature is 2~K.}
\label{fig:PL}
\end{figure}

Photoluminescence (PL) spectra of the sample for different magnetic fields $B_\text{F}$ applied along the sample normal (in Faraday geometry) are shown in Fig.~\ref{fig:PL}(a). At $B_\text{F} = 0$~T the PL spectrum spreads from 1.5185 to 1.5225~eV. The width of the spectrum allows to estimate the Fermi energy of the 2DEG, $E_\text{F} \approx 4$~meV, which corresponds to the electron concentration of $n_\text{e} = m_\text{e} E_\text{F} / \pi \hbar^2 \approx 1.1 \times 10^{11}$~cm$^{-2}$, where $m_\text{e} = 0.067 m_0$ is the electron effective mass and $m_0$ is the free electron mass. With increasing magnetic field, distinct PL peaks appear corresponding to the discrete energy levels due to Landau quantization in the two-dimensional system. From Fig.~\ref{fig:PL}(b) it is seen that the peak energy positions increase linearly with the field, following the dependence $E_n=E_0 + \hbar \omega_\text{c} (n+1/2)$, where $E_0 \approx 1.5185$~eV corresponds to the gap between the conduction and valence bands in the QW, $\omega_\text{c} = eB/m_\text{e}$ is the cyclotron precession frequency. The experiment gives $\hbar \omega_\text{c} \approx 1.63$~meV/T, in good agreement with the calculated electron cyclotron energy of 1.73~meV/T. This agreement indicates that free electrons recombine with bound holes. Interestingly, it was reported in Ref.~\cite{Koudinov2016} that for quasiresonant excitation of the 2DEG the electrons recombine with mobile holes which results in the separation between the PL lines of 2.1~meV/T equal to the sum of the electron and hole cyclotron energies. 

\begin{figure*}
\includegraphics[width=1.9\columnwidth]{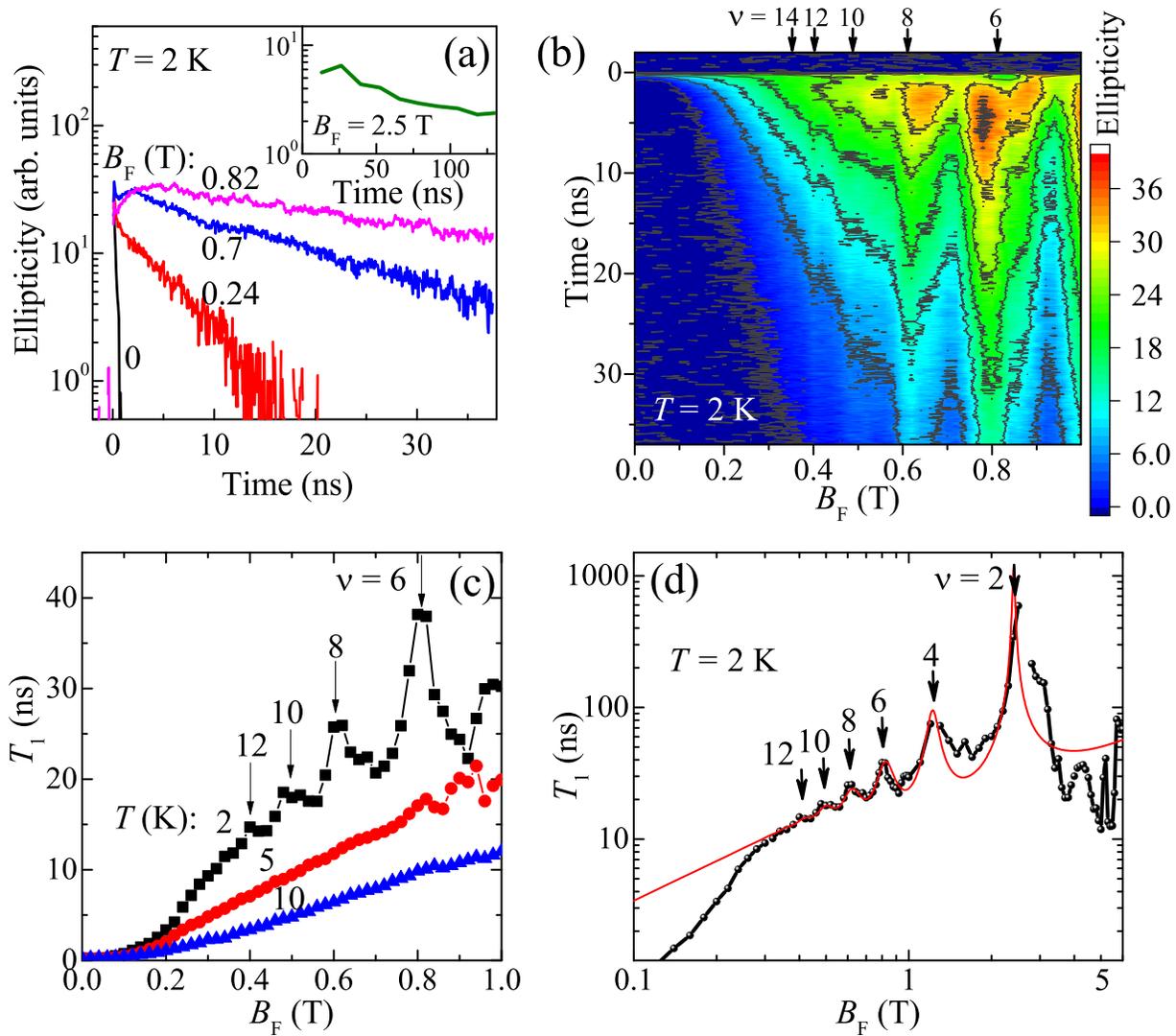}
\caption{Spin dynamics in Faraday geometry. (a) Dynamics of the Kerr ellipticity for different magnetic fields applied in Faraday geometry. Inset shows the long-term dynamics close to filling factor $\nu = 2$ (see text). $T = 2$~K. (b) Two-dimensional plot showing the ellipticity as a function of time and magnetic field. $T = 2$~K. (c) Magnetic field dependence of the longitudinal spin relaxation time $T_1$ at different temperatures. (d) Magnetic field dependence of $T_1$ at $T = 2$~K in a large range of magnetic fields. The red line shows the theoretical fit to the experimental data.}
\label{fig:T1vsBT}
\end{figure*}

In what follows we examine the dynamics of the Kerr ellipticity. In these experiments the photon energy of both the pump and probe beams is 1.523~eV, close to the Fermi energy of the 2DEG. The ellipticity dynamics is shown in Fig.~\ref{fig:T1vsBT}(a) for different values of the magnetic field $B_\text{F}$ applied in Faraday geometry. The comprehensive picture of the ellipticity dynamics as a function of the magnetic field is shown by the color map in Fig.~\ref{fig:T1vsBT}(b). During the first $\sim 5$~ns (at $B_\text{F} \gtrsim 0.5$~T) the dynamics shows indications of being  nonexponential and even nonmonotonous. At later times, the decay is close to being exponential, allowing us to determine the longitudinal spin relaxation time $T_1$. The dependence of $T_1$ on magnetic field is shown in Fig.~\ref{fig:T1vsBT}(c) for three different temperatures and in Fig.~\ref{fig:T1vsBT}(d) for $T = 2$~K over a wide range of magnetic fields up to 6~T. The following conclusions can be derived from these dependencies.
(i) At low $B_\text{F}\lesssim 0.3$~T the relaxation time $T_1 \propto B_\text{F}^2$.
(ii) At higher $B_\text{F}\gtrsim 0.3$~T , $T_1$ increases almost linearly with $B_\text{F}$. The rate of this increase $dT_1/dB_\text{F}$ is roughly inversely proportional to the temperature $T$. So we can conclude that at $B_\text{F} \gtrsim 0.3$~T the time $T_1 \propto B_\text{F} / T$.
(iii) For $B_\text{F} \gtrsim 0.5$~T and at the lowest $T = 2$~K, 
distinct peaks in the dependence $T_1(B_\text{F})$ appear. The positions of these peaks correspond to the even filling factors $\nu$, i.e. where the $\nu / 2$ Landau levels are completely filled by electrons with spins both parallel and antiparallel to the magnetic field. Around $\nu = 2$ ($B_\text{F} \approx 2.5$~T) the ellipticity dynamics is long-lived and nonexponential [inset in Fig.~\ref{fig:T1vsBT}(a)] with a small amplitude, which makes $T_1$ difficult to determine. Obviously, the peak contrast increases for smaller $\nu$. We do not observe any dependence of the $T_1$ peak positions on the laser photon energy, which mostly affects the magnitude of the Kerr ellipticity signal.

\begin{figure*}
\includegraphics[width=1.9\columnwidth]{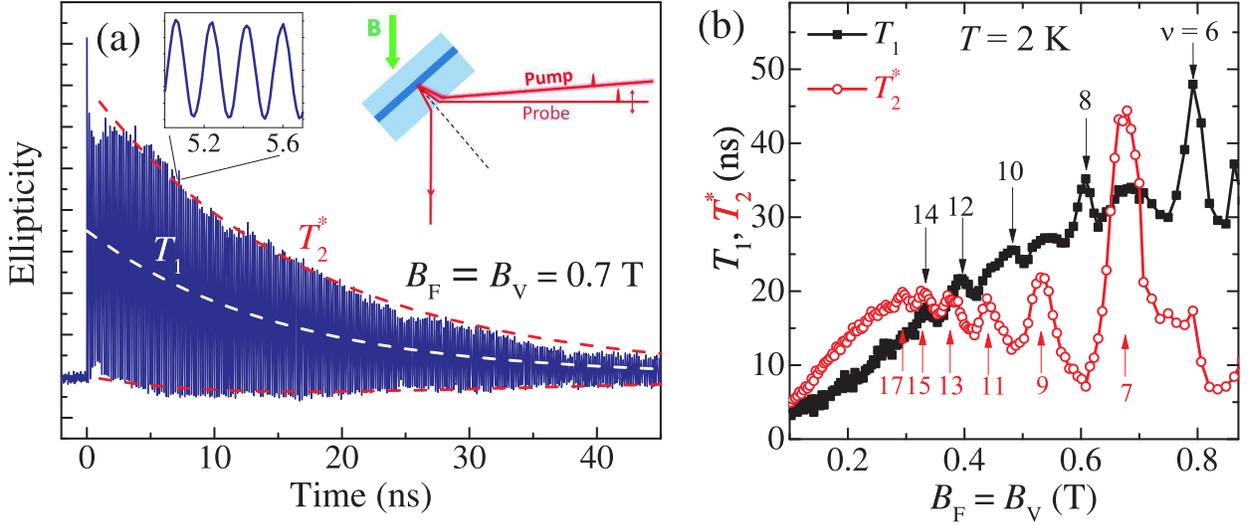}
\caption{Spin dynamics for the sample tilted by $45^o$ relative to the magnetic field direction. (a) Kerr ellipticity dynamics for longitudinal and transverse magnetic field components of 0.7~T. Inset shows a closeup of the dynamics (left) and the experimental geometry (right). $T = 5$~K. (b) Dependence of the longitudinal spin relaxation time $T_1$ and transverse spin dephasing time $T_2^*$ on $B_\text{F} = B_\text{V} = B/\sqrt{2}$. $T = 2$~K.}
\label{fig:T1T2vsB}
\end{figure*}

Next we switch to the $45^\circ$ geometry, where the sample normal is tilted by $\sim 45^\circ$ with respect to the magnetic field and the pump and probe beams [see inset in Fig.~\ref{fig:T1T2vsB}(a)]. The incident beams are refracted at the sample surface and arrive to the QW at the angle of $\sim 12^\circ$ relative to the sample normal. Thus, the pump pulse creates spin polarization having close components along and perpendicular to the magnetic field, $S_\parallel \approx S_\perp$. While $S_\parallel (t)$ monotonically decays with time $T_1$, $S_\perp (t)$ precesses about the magnetic field and decays with time $T_2^*$. In this way the spin projection onto the sample normal, which is finally detected by the probe beam, is \cite{Belykh2016a}:
\begin{equation}
S(t) = \frac{S_\parallel}{\sqrt{2}} \exp\left(-\frac{t}{T_1}\right) + \frac{S_\perp}{\sqrt{2}} \exp\left(-\frac{t}{T_2^*}\right)\cos(\Omega_\text{L}t),
\label{eq:St}
\end{equation}
where $\Omega_\text{L} = g\mu_\text{B}B/\hbar$ is the Larmor frequency, $\mu_\text{B}$ is the Bohr magneton and $g$ is the electron $g$ factor. Thus both $T_1$ and $T_2^*$ can be determined in this experiment.

The measured ellipticity dynamics is shown in Fig.~\ref{fig:T1T2vsB}(a). According to Eq.~\eqref{eq:St}, the decay of the precession amplitude gives $T_2^*$, while the decay of the nonprecessing component gives $T_1$. From the fit of Eq.~\eqref{eq:St} to the experimental data we also find $|g| = 0.40$, while the sign of $g$ factor is expected to be negative for the transition energy of the studied QWs \cite{Yugova2007Univ} which is linked with the $g$ factor via the Roth equation \cite{Roth1959}. So that $g = -0.40$. Figure~\ref{fig:T1T2vsB}(b) shows the dependence of $T_1$ and $T_2^*$ on the Faraday magnetic field component $B_\text{F}$, which is equal to the Voigt component along the sample surface $B_\text{V} = B_\text{F} = B/\sqrt{2}$. The magnetic field dependence of $T_1$ is expectedly similar to that in pure Faraday geometry (Fig.~\ref{fig:T1vsBT}) with sharp peaks at even $\nu$. The transverse spin relaxation time $T_2^*$ monotonically increases with $B_\text{F} = B_\text{V}$ for $B_\text{F} \lesssim 0.3$~T. Interestingly, in this magnetic field range the limiting relation $T_2^* \approx 2 T_1$ holds, so that the transverse spin relaxation is limited by the longitudinal spin relaxation. At higher magnetic fields distinct peaks in the dependence $T_2^*(B_\text{F})$ appear at the positions corresponding to the odd filling factors, in agreement with works \cite{Fukuoka2008,Fukuoka2010,Larionov2015,Larionov2017,Larionov2020}.

\section{Theory of the longitudinal spin relaxation}

\subsection{Spin precession and nutation}

The dynamics of the electron spin between consecutive collisions is governed by the Bloch equation without damping,
\begin{equation}
\label{spinBlochEquation}
\frac{d\mathbf{s}}{dt}=\mathbf{\Omega}\times\mathbf{s},
\end{equation}
which means that at every time the spin momentarily rotates about the unitary axis parallel to $\mathbf{\Omega}$ with angular frequency $\Omega$. There are two contributions to the angular frequency,
\begin{equation}
\label{Omega}
\mathbf{\Omega}=\mathbf{\Omega}_\textrm{L}+\mathbf{\Omega}_\textrm{D}.
\end{equation}
The former contribution is due to the Zeeman splitting in external magnetic field, $\mathbf{\Omega}_\textrm{L}=\Omega_\textrm{L}\mathbf{e}_z$ with $\hbar\Omega_\textrm{L}=g\mu_\textrm{B}B \sim-0.02\text{ meV}\,(B/1\text{ T})$ (note, here we consider only the Faraday configuration). The latter in-plane contribution is due to the Dresselhaus spin-orbit interaction in zinc-blende-type semiconductors \cite{Dresselhaus1955,DyakonovPerel1971,DyakonovKachorovskii1986}, $\hbar\mathbf{\Omega}_\textrm{D}=2\beta_\textrm{D}(-k_x\mathbf{e}_x+k_y\mathbf{e}_y)$ with $\beta_\textrm{D}=\gamma\pi^2/a^2\sim 0.4\text{ meV}\,\text{nm}$, where the QW width $a\sim25$~nm, $\gamma\sim28\text{ meV}\,\text{nm}^3$ \cite{MalcherEtal1986,Winkler2003}, note the existing uncertainty in determining $\gamma$ \cite{ChantisEtal2006,KrichHalperin2007,WalserEtal2012}). Thus, $\hbar\Omega_\textrm{D}=2\beta_\textrm{D}k_\textrm{F}\sim0.07$~meV, where $k_\textrm{F} = (2\pi n_\text{e})^{1/2}$ is the Fermi wave vector. We consider here only the case of the Dresselhaus interaction, but generalization to the case of the Rashba interaction \cite{Vasko1979,BychkovRashba1984}, which can be important in the case of an asymmetric QW \cite{EldridgeEtal2010,EnglishEtal2013}, is straightforward.

By the Euler theorem the spin orientation at some later time can be obtained from the initial orientation at zero time by rotation around a unitary axis $\boldsymbol{\zeta}$ through an angle $\alpha$, and the axis orientation and angle generally depend on time, $\boldsymbol{\zeta}=\boldsymbol{\zeta}(t)$ and $\alpha=\alpha(t)$. The orientation can be specified by a quaternion
\begin{equation}
\label{quaternionLambda}
\Lambda=\cos\frac{\alpha}{2}+\boldsymbol{\zeta}\sin\frac{\alpha}{2}=\exp\left(\frac{\boldsymbol{\zeta}\alpha}{2}\right),
\end{equation}
which corresponds to the above rotation and is also time dependent, $\Lambda=\Lambda(t)$. Quaternions are a special type of hypercomplex numbers used in various physical problems and particularly effective in the study of rotational dynamics \cite{Sobyanin2016,GorhamLaughlin2019,KadekEtal2019,GubbiottiEtal2019}. The knowledge of the time behavior of $\Lambda$ provides the knowledge of the spin dynamics.

Prior to finding $\Lambda$, let us recall some necessary facts from the theory of quaternions \cite{BranetsShmyglevskii1973,Zhuravlev2001,Girard2007}. A quaternion $\mathrm{M}=\mu_0+\boldsymbol{\mu}$ is the sum of a scalar part $\mu_0$ and a vector part $\boldsymbol{\mu}$. The conjugate quaternion is $\bar{\mathrm{M}}=\mu_0-\boldsymbol{\mu}$; for the quaternion \eqref{quaternionLambda} we have $\bar{\Lambda}=\cos(\alpha/2)-\boldsymbol{\zeta}\sin(\alpha/2)=\exp\left(-\boldsymbol{\zeta}\alpha/2\right)$. Any vector is a quaternion with zero scalar part, e.g., the radius vector $\mathbf{r}$ can be considered as the quaternion $0+\mathbf{r}$. The product of two quaternions $\mathrm{M}$ and $\mathrm{N}=\nu_0+\boldsymbol{\nu}$ is defined as
\begin{equation}
\label{quaternionProduct}
\mathrm{M}\circ \mathrm{N}=\mu_0\nu_0-\boldsymbol{\mu}\cdot\boldsymbol{\nu}+\mu_0\boldsymbol{\nu}+\nu_0\boldsymbol{\mu}+\boldsymbol{\mu}\times\boldsymbol{\nu}.
\end{equation}
The product is associative, $(\mathrm{\Lambda}\circ\mathrm{M})\circ\mathrm{N}=\mathrm{\Lambda}\circ(\mathrm{M}\circ\mathrm{N})=\mathrm{\Lambda}\circ\mathrm{M}\circ\mathrm{N}$, but not commutative, $\mathrm{M}\circ\mathrm{N}\neq\mathrm{N}\circ\mathrm{M}$. The latter property reflects that two successive rotations of a spin generally give another final orientation than the same rotations applied in inverse order. For the product \eqref{quaternionProduct} the conjugation property $\overline{\mathrm{M}\circ\mathrm{N}}=\bar{\mathrm{N}}\circ\bar{\mathrm{M}}$ is valid. Every quaternion of the form \eqref{quaternionLambda} has unitary norm $\Lambda\circ\bar{\Lambda}=1$.

The momentary spin $\mathbf{s}=\mathbf{s}(t)$ is obtained from the initial spin $\mathbf{s}_0=\mathbf{s}(0)$ by the rotation specified by $\Lambda$ and thus is expressed by:
\begin{equation}
\label{spinRotation}
\mathbf{s}=\Lambda\circ\mathbf{s}_0\circ\bar{\Lambda}.
\end{equation}
The angular velocity is related to the quaternion via
\begin{equation}
\label{OmegaViaLambda}
\mathbf{\Omega}=2\frac{d\Lambda}{dt}\circ\bar{\Lambda}.
\end{equation}
Interestingly, every quaternion, $\mathrm{M}$ say, can be written in an alternative form using the spin matrix basis, $\mathrm{M}=\mu_0\sigma_0-i\boldsymbol{\mu}\cdot\boldsymbol{\sigma}$, where $\sigma_0$ is the identity matrix and $\boldsymbol{\sigma}$ is the vector of the usual Pauli matrices.

Now we can calculate $\Lambda$ from Eq.~\eqref{OmegaViaLambda}. An electron in a magnetic field executes a cyclotron motion with angular velocity $\boldsymbol{\omega}_\textrm{c}=\omega_\textrm{c}\mathbf{e}_z$, and this motion is described by a quaternion $\mathrm{M}=\exp\left(\boldsymbol{\omega}_\textrm{c}t/2\right)$. The Dresselhaus angular velocity $\mathbf{\Omega}_\text{D}$ and the angular velocity $\mathbf{\Omega}$ rotate in inverse direction, and this rotation is described by the conjugate quaternion~$\bar{\mathrm{M}}$. We seek the solution as a product of the latter quaternion and another unknown quaternion~$\mathrm{N}$, so that $\Lambda=\bar{\mathrm{M}}\circ\mathrm{N}$. Substituting this quaternionic product into Eq.~\eqref{OmegaViaLambda} and using the aforementioned properties and definitions, one finds that the angular velocity corresponding to an extra rotation given by the quaternion $\mathrm{N}$ is a constant vector
\begin{equation}
\label{omega0}
\boldsymbol{\omega}_0=\boldsymbol{\Omega}_0+\boldsymbol{\omega}_\textrm{c},
\end{equation}
where $\boldsymbol{\Omega}_0=\boldsymbol{\Omega}(0)$ is the angular velocity \eqref{Omega} of spin rotation taken at zero time. This vector has the absolute value
\begin{equation}
\label{omega0absoluteValue}
\omega_0=\sqrt{(\omega_\textrm{c}+\Omega_\textrm{L})^2+\Omega_\textrm{D}^2}
\end{equation}
and for the experimental conditions $\omega_0\approx\omega_\textrm{c}$ because typically $\omega_\textrm{c}$ greatly exceeds $\Omega_\textrm{L}$ and~$\Omega_\textrm{D}$. Therefore, $\mathrm{N}=\exp\left(\boldsymbol{\omega}_0t/2\right)$.

Thus, the quaternion describing spin orientation at time $t$ is given by
\begin{equation}
\label{LambdaSolution}
\Lambda=\exp\left(\boldsymbol{-\omega}_\textrm{c}t/2\right) \circ \exp\left(\boldsymbol{\omega}_0 t/2\right).
\end{equation}
Taking into account Eq.~\eqref{spinRotation}, we see that the final spin orientation is obtained from the initial orientation by superposition of two successive rotations around fixed axes, the first around $\mathbf{e}_0=\boldsymbol{\omega}_0/\omega_0$ through angle $\omega_0 t$ and the second around $\mathbf{e}_z$ through angle $-\omega_\textrm{c}t$ (i.e., the rotation is made in reverse direction). Equation \eqref{LambdaSolution} means that the same result can be obtained by directly transforming Eq.~\eqref{spinBlochEquation} to the rotating frame of reference, whose rotation is described by~$\bar{\mathrm{M}}$, and then dealing with the modified but nonrotating angular velocity, which determines~$\mathrm{N}$.

The forward rotation of the spin occurs in an inclined plane and brings about a change in the vertical ($z$-axis) spin projection. On the other hand, the backward rotation occurs in the horizontal ($xy$) plane and has no influence on the spin projection resulting from the forward rotation, but instead returns the spin for small times to almost the same azimuthal orientation as the one at zero time. Since $\omega_0\approx\omega_\textrm{c}$, the total azimuthal phase changes slowly. The polar angle returns to its initial position in time $\tau_0=2\pi/\omega_0$ and then the azimuthal angle changes by $\Delta\phi=2\pi-\omega_\textrm{c}\tau_0=2\pi(\omega_0-\omega_\textrm{c})/\omega_0\ll1$. This means that the spin executes a slow precession around $\mathbf{e}_z$ with angular frequency $\Omega_\textrm{prec}=\omega_0-\omega_\textrm{c}$ and a simultaneous fast nutation with frequency $\omega_0$.

\subsection{Spin relaxation}

From the above considerations it follows that only the first rotation given by the quaternion $\mathrm{N}$ determines a change in the vertical spin component and contributes to the longitudinal spin relaxation. Naturally, such a change is solely due to the inclined $\mathbf{e}_0$, i.e., due to a nonzero $\Omega_\textrm{D}$. However, the polar angle, which characterizes the inclination of the spin with respect to $\mathbf{e}_z$, oscillates with frequency $\omega_0$, but remains near its initial inclination in a small phase range equal to the $\mathbf{e}_0$ inclination angle, $\delta\theta\sim\Omega_\textrm{D}/\omega_0\ll1$. This means that the further spin relaxation is impossible without a change in the rotational phase of the total angular velocity $\mathbf{\Omega}$ and hence in the rotational phase of the electron. This change requires collisions, which, quite importantly, can be of any nature and are not necessarily contributing to the sample mobility \cite{WuNing2000,GlazovIvchenko2002}.

In the experiment the time $\tau_\textrm{col}$ between two successive collisions exceeds greatly the period $\tau_\textrm{c}\sim2\text{ ps}\,(B/1\text{ T})^{-1}$ of cyclotron motion and hence the period of spin nutation, $\omega_0\tau_\textrm{col}\gg1$. Then every collision causes a random polar phase shift $\sim\delta\theta$ independent of $\tau_\textrm{col}$ as well as having random sign and adds a phase variance $\sim\delta\theta^2$ to the momentary variance. After the time $T_1$ of longitudinal spin relaxation the electron has executed $n=T_1/\tau_\textrm{col}$ collisions and the total phase change is of the order of unity, $n\delta\theta^2\sim1$. The spin relaxation time becomes
\begin{equation}
\label{T1}
T_1\sim\tau_\textrm{col}\Bigl(\frac{\omega_0}{\Omega_\textrm{D}}\Bigr)^2.
\end{equation}
The linear dependence on the collision time is inverse to the $1/\tau_\textrm{col}$ law expected for the usual Dyakonov-Perel mechanism and mimics the Elliot-Yafet mechanism \cite{Elliott1954,Yafet1963}. Nonetheless, it does not require a spin-flip upon collision, but simply reflects the influence of the external magnetic field and is consistent with previous results \cite{Ivchenko1973,BurkovBalents2004,Glazov2004,FabianEtal2007,ZhouEtal2013}.

Note that though Eq.~\eqref{T1} formally coincides with the high-field limit of the modified Dyakonov-Perel behavior $T_1 \sim (1+\omega_\textrm{c}^2\tau_\textrm{col}^2)/\Omega_\textrm{D}^2\tau_\textrm{col}$ \cite{Ivchenko1973}, it is more general and is valid not only in the strong-scattering regime for the electron spin itself, when $\Omega_\textrm{D}\tau_\textrm{col}\ll1$, but also in the weak-scattering regime, when $\Omega_\textrm{D}\tau_\textrm{col}\gg1$. In our case Eq.~\eqref{T1} is not a consequence of that behavior because $\Omega_\textrm{D}\tau_\textrm{col}\gg1$ in the experiment and the standard zero-field estimation $T_1\sim1/\Omega_\textrm{D}^2\tau_\textrm{col}$ does not apply.

\subsection{Relation between spin relaxation and spatial diffusion}

Besides a polar phase shift for the electron spin, every collision brings about a corresponding shift in the electron position because both the shifts have the same reason, the change in the rotational phase of the electron induced by the collision. This means that the spin diffusion on the Bloch sphere leading to spin relaxation and the electron diffusion in space are closely related. Since $\omega_\textrm{c}\tau_\textrm{col}\gg1$, the electron makes many gyrations in the cyclotron orbit between two successive collisions and its spatial shift upon a collision is on the order of the cyclotron radius $r_\textrm{c}=v_\textrm{F}/\omega_\textrm{c}$, where $v_\textrm{F}$ is the Fermi velocity. The spatial diffusion coefficient for the electron is \cite{ArtsimovichSagdeev1979,Laikhtman1994,HelanderSigmar2002}
\begin{equation}
\label{highFieldD}
D\sim\frac{r_\textrm{c}^2}{\tau_\textrm{col}},
\end{equation}
while the spin polar-angle diffusion coefficient is
\begin{equation}
D_\text{s} \sim \frac{\delta\theta^2}{\tau_\textrm{col}} \sim \frac{1}{\tau_\textrm{col}} \left(\frac{\Omega_\textrm{D}}{\omega_0}\right)^2  \sim \frac{1}{T_1}.
\label{eq:Ds}
\end{equation}
The ratio of the first to the second coefficient has the dimension of a squared length,
\begin{equation}
l_\textrm{s}^2=\frac{D}{D_\textrm{s}},
\label{eq:ls}
\end{equation}
so that $l_\textrm{s}$ is the spin diffusion length equal to the spatial shift of the electron (more exactly, of its gyrocenter) in the spin relaxation time. $l_\textrm{s} \sim v_\textrm{F}/\Omega_\textrm{D}=\hbar^2/2m_\textrm{e}\beta_\textrm{D} \sim 1.3$ $\mu$m depends on the sample characteristics only. The expression for $l_\textrm{s}$ generalizes a recent result obtained for the Rashba interaction in polycrystalline graphene in zero magnetic field and in the strong-scattering regime for the spin, where $\Omega_\textrm{R}\tau_\textrm{col}\ll1$ and $T_1\sim1/\Omega_\textrm{R}^2\tau_\textrm{col}$ \cite{CummingsEtal2019}, and appears to be valid in magnetic fields satisfying $\omega_\textrm{c}\tau_\textrm{col}\gg1$, when the constraints on $\Omega_\textrm{D(R)}\tau_\textrm{col}$ can be lifted. We get from Eqs.~\eqref{eq:Ds}~and~\eqref{eq:ls}
\begin{equation}
\label{T1viaD}
T_1\sim\frac{l_\textrm{s}^2}D.
\end{equation}
Thus, the spin relaxation time appears to be inversely proportional to the electron diffusion coefficient.

\section{Discussion}

Figure \ref{fig:T1vsBT}(d) shows the $B$ dependence of $T_1$ at $T = 2$~K in the magnetic field range up to 6~T. At low fields $T_1\propto B^2$ and at high fields $T_1\propto B$ with a crossover at $B\sim0.3$~T, which according to Eq.~\eqref{T1viaD} means that the diffusion coefficient changes its behavior from $D\propto B^{-2}$ to $D\propto B^{-1}$ as $B$ increases. The low-field behavior corresponds to the well-known classical diffusion of a plasma across magnetic field~\cite{HelanderSigmar2002}, while the high-field behavior is anomalous. The study of the anomalous plasma diffusion in a magnetic field goes back to 1949, when it was first experimentally found that, contrary to the expected classical behavior, the plasma diffusion occurs much faster and reduces the impact of the magnetic field on the plasma confinement \cite{Bohm1949}. The corresponding coefficient of anomalous diffusion, called the Bohm diffusion, is given by
\begin{equation}
\label{BohmDiffusionCoefficient}
D_\textrm{B}\sim\frac{k_\textrm{B}T}{eB}.
\end{equation}
There is no unique explanation for the Bohm diffusion as well as for other types of deviations from the classical behavior. Such departures are usually related to specifics of electromagnetic fluctuations, turbulence, strong coupling, and transition from a 3D to a 2D situation \cite{Spitzer1960,Braginskii1965,TaylorMcNamara1971,MarchettiEtal1984,OttBonitz2011,FengEtal2014}. Beyond terrestrial laboratories, anomalous plasma behavior also reveals itself in `cosmic laboratories' such as rotating neutron stars, whose magnetic fields reach $10^8-10^{11}$~T and make them efficient generators of dense relativistic plasmas \cite{IstominSobyanin2007,IstominSobyanin2011}: recent observations and simulations show the role of Bohm diffusion in the transport of high-energy particles in their surroundings \cite{PorthEtal2016,LiuEtal2019}.

In the experiment we observe $T_1\propto B/T$ at $B \gtrsim 0.3$~T [Fig.~\ref{fig:T1vsBT}(c)], which, together with Eq.~\eqref{T1viaD}, allows us to conclude that the diffusion coefficient behaves similarly to the Bohm diffusion coefficient. After substituting Eq.~\eqref{BohmDiffusionCoefficient} into Eq.~\eqref{T1viaD}, we arrive at a spin relaxation time of the form
\begin{equation}
\label{T1ForBohmDiffusion}
T_1\sim\tau_0\frac{\hbar\omega_\textrm{c}}{k_\textrm{B}T}\frac{\rho_0}{\rho(\varepsilon_\textrm{F})},
\end{equation}
where the effective time $\tau_0\sim m_\textrm{e} l_\textrm{s}^2/\hbar\sim1$~ns, similarly to the spin diffusion length, depends on the sample characteristics only. In Eq.~\eqref{T1ForBohmDiffusion} we have considered the dependence of the collision time on the density of states $\rho(\varepsilon_\textrm{F})$ at the Fermi level, $\tau_\textrm{col}=\tau_{\textrm{col}\,0}\rho_0/\rho(\varepsilon_\textrm{F})$, where $\tau_{\textrm{col}\,0}$ and $\rho_0=m_\textrm{e}/\pi\hbar^2$ are the collision time and the density of states at low magnetic fields \cite{BurkovBalents2004}; by Eq.~\eqref{T1} an analogous relation is valid for the spin relaxation times. This relation reflects the proportionality of the transition probability to the density of final states \cite{Lundstrom2000}. For every single Landau level in $\rho(\varepsilon)$ we take a Gaussian form of broadening with the standard deviation $\sigma=\sqrt{\sigma_0^2+\sigma_\textrm{th}^2}$. Here $\sigma_0\propto\sqrt{B}$ \cite{AndoUemura1974,RaikhShahbazyan1993} while $\sigma_\textrm{th}\sim k_\textrm{B}T$ takes account of the finite temperature and follows from the idea that a sample at a low, but nonzero temperature, can approximately be considered equivalent to a set of samples with Fermi energies distributed about the Fermi energy of the initial sample with the probability density $-\partial f/\partial\varepsilon$, where $f$ is the usual Fermi-Dirac distribution \cite{Shoenberg1984}.

Figure \ref{fig:T1vsBT}(d) demonstrates the comparison of the theoretical dependence \eqref{T1ForBohmDiffusion} and the experimental results. Good coincidence is seen in the range of magnetic fields $B = B_\text{F} >0.3$~T. The oscillatory $B$ dependence of $T_1$, where the maxima of $T_1$ correspond to even $\nu$ and the minima to odd $\nu$, appears because of the oscillatory dependence of $\rho(\varepsilon_\textrm{F})$: at low fields the density of states reduces to the constant $\rho_0$, but at high fields, when the separation between adjacent Landau levels becomes greater than the width of individual levels, oscillations occur. In this sense the effect is similar to the Shubnikov-de-Haas and quantum Hall effects \cite{Datta1995}. We want to stress that the proportionality factor in the linear dependence between $B$ and $T_1$ at low fields, which determines the overall inclined trend of~\eqref{T1ForBohmDiffusion}, is not a free parameter of the model but is related to the sample characteristics and temperature after adopting the diffusion coefficient of the form~\eqref{BohmDiffusionCoefficient}.

Thus, the observed behavior of the spin relaxation time at magnetic fields above 0.3~T can be a signature of anomalous Bohm diffusion of the two-dimensional electron plasma, and below 0.3~T the anomalous diffusion turns to the usual classical diffusion. The observation of classical behavior at low fields and anomalous behavior at high fields is consistent with earlier plasma results \cite{OkudaDawson1973,MarchettiEtal1984}.

In the usual nondegenerate thermal plasma, the anomalous diffusion can loosely be interpreted as the maximum attainable diffusion \cite{Taylor1961}. In the weak-scattering regime, the diffusion coefficient $D$ is given by Eq.~\eqref{highFieldD} in which the cyclotron radius $r_\textrm{c}=v_\textrm{th}/\omega_\textrm{c}$ is determined by the thermal velocity $v_\textrm{th}\sim\sqrt{k_\textrm{B}T/m_\textrm{e}}$. In the strong-scattering regime, on the other hand,  $D\sim v_\textrm{th}^2\tau_\textrm{col}$, and the diffusion rate reaches its maximum at the formal anomalous collision time $\tau_\textrm{a}\sim1/\omega_\textrm{c}$ corresponding to a crossover between the two scattering regimes and for which $D\sim v_\textrm{th}^2/\omega_\textrm{c}$ becomes equal to the Bohm value~\eqref{BohmDiffusionCoefficient}. However, we deal with a strongly degenerate electron gas, and it is interesting to estimate the anomalous collision time $\tau_{\textrm{a}\,\textrm{deg}}$ in this case. From the equality $D_\textrm{B}\rho(\varepsilon_\textrm{F})/\rho_0\sim r_\textrm{c}^2/\tau_{\textrm{a}\,\textrm{deg}}$ we obtain
$\omega_\textrm{c}\tau_{\textrm{a}\,\textrm{deg}}\sim2n_\textrm{s}/\rho(\varepsilon_\textrm{F})k_\textrm{B}T$.
On the right-hand side we see the ratio of the total electron number to their number involved in the temperature smearing of the Fermi-Dirac distribution near the Fermi energy, so that $\omega_\textrm{c}\tau_{\textrm{a}\,\textrm{deg}}\gg1$ and, in contrast to the nondegenerate situation, Bohm diffusion of the degenerate electron gas happens in the weak-scattering regime. Seemingly, the inequality $\tau_{\textrm{a}\,\textrm{deg}}\gg\tau_\textrm{a}$ reflects inhibition of the single-electron diffusion as a result of a decrease in the probability of individual jumps because of effects of quantum degeneracy. An analogous situation was observed earlier in studies of the spin-polarized transport in the regime of Pauli blockade \cite{CadizEtal2013,CadizEtal2015}.

It is instructive to compare the longitudinal spin relaxation in 2DEG and bulk systems. In $n$-doped bulk GaAs containing a low-mobility electron gas in the metallic phase, the spin relaxation is governed by the classical Dyakonov-Perel mechanism with frequent electron collisions. In this regime $T_1 \propto 1/\tau_\text{col}$, contrary to Eq.~\eqref{T1}. The dependence of $T_1$ on $B$ is not strong, for $B \lesssim 1$~T it is governed by the weak localization, and only for higher $B$, $T_1$ increases due to cyclotron electron precession \cite{Belykh2018}. Interestingly, in $n$-GaAs with low electron concentration, so that the electrons are localized at low $T$, the behavior of $T_1$ with $B$ is similar to that we observe in the 2DEG, where the electron localization is induced and controlled by the magnetic field. The time $T_1$ increases linearly with $B$ with the slope decreasing with temperature \cite{Belykh2017}.  This behaviour was explained by spin diffusion towards optimal pairs of charged and neutral donors \cite{Kavokin08}, while the inverse (anomalous) dependence of the diffusion coefficient with $B$ reflects the decrease of the spin diffusion efficiency with increasing spread of the Larmor frequencies $\delta \Omega_\text{L}=\delta g \mu_\text{B} B /\hbar$.

\section{Conclusion}
We have studied the longitudinal and transverse spin dynamics in a high-mobility 2DEG confined in a GaAs/AlGaAs quantum well at low temperatures and high magnetic fields, using the extended pump-probe Kerr rotation technique. A magnetic field applied along the sample normal drastically suppresses both the longitudinal and transverse spin relaxation. As the magnetic field increases, the initially quadratic magnetic-field dependence of the spin relaxation time becomes linear, with a slope inversely proportional to the temperature. At higher fields  and low temperatures, the magnetic field dependence of the spin relaxation times eventually takes on an oscillatory character, so that $T_1$ and $T_2^*$ have maxima at even and odd filling factors, respectively. Using quaternions, we theoretically show that the observed strong magnetic damping of the spin relaxation is related to limited spin nutation appearing because of electron gyration in a magnetic field that causes a rotation of the spin-orbit field. The spin relaxation appears to be closely related to the electron spatial diffusion, and it is theoretically expected that the longitudinal spin relaxation time is inversely proportional to the spatial diffusion coefficient. The transition from the quadratic to the linear magnetic-field dependence of the spin relaxation time can be related to a transition from the classical to the Bohm diffusion of the electrons and reflects an anomalous behavior of the 2DEG in a magnetic field, analogous to that observed in magnetized plasmas. The oscillations of $T_1$ with magnetic field are related to the oscillations of the density of states at the Fermi level and correspond to the transition at high fields and low temperatures to the Shubnikov-de-Haas and quantum Hall regime.

\section{Acknowledgments}
\begin{acknowledgments}
Samples similar to the ones considered here were investigated by B.~Ashkinadze et al. \cite{Ashkinadze2001}, and we are grateful to him for providing  the heterostructure and useful discussions. We are grateful to E. Evers, A.~V.~Larionov, E. Stepanets-Khussein, and V.~F.~Sapega for valuable advices. The work was supported by the Deutsche Forschungsgemeinschaft in the frame of the ICRC TRR 160 (project A1). V.V.B, M.V.K. and D.N.S. acknowledge the financial support of the Russian Science Foundation through the grant No.~18-72-10073.
\end{acknowledgments}

\end{document}